\documentclass{elsart}
\usepackage{graphicx}

\usepackage{MnSymbol} %REEMPLAZAR!!
\usepackage{amsmath}
\usepackage{color}
\usepackage[dvipsnames]{xcolor}
\usepackage{bm}% bold math

\usepackage{cite}

\journal{Nuclear Physics A}

\begin{document}
\begin{frontmatter}
\title{Pairing in the BCS and LN approximations using continuum single particle level density}
% title[Pairing in the BCS and LN approximations using continuum density]{Pairing in the BCS and LN approximations using continuum single particle level density}
\author{R. M. Id Betan$^{1,2,3}$, C. E. Repetto$^{1,2}$}
\address{$^{1}$Instituto de F\'isica Rosario (CONICET-UNR), Bv. 27 de Febrero 210 bis, S2000EZP Rosario. Argentina.}
\address{$^{2}$Facultad de Ciencias Exactas, Ingenier\'ia y Agrimensura (UNR), Av. Pellegrini 250, S2000BTP Rosario. Argentina.}
\address{$^{3}$Instituto de Estudios Nucleares y Radiaciones Ionizantes (UNR), Riobamba y Berutti, S2000EKA Rosario. Argentina.}

\date{\today}

\begin{abstract}
Understanding the properties of drip line nuclei requires to take into account the correlations with the continuum spectrum of energy of the system. This paper has the purpose to show that the continuum single particle level density is a convenient way to consider the pairing correlation in the continuum. Isospin mean-field and isospin pairing strength are used to find the Bardeen-Cooper-Schrieffer (BCS) and Lipkin-Nogami (LN) approximate solutions of the pairing Hamiltonian. Several physical properties of the whole chain of the Tin isotope, as gap parameter, Fermi level, binding energy, and one- and two-neutron separation energies, were calculated and compared with other methods and with experimental data when they exist. It is shown that the use of the continuum single particle level density is an economical way to include explicitly the correlations with the continuum spectrum of energy in large scale mass calculation. It is also shown that the computed properties are in good agreement with experimental data and with more sophisticated treatment of the pairing interaction.
\end{abstract}

\begin{keyword}
Properties of nuclei                  \sep Single particle density \sep Pairing
\PACS 21.10.-k  \sep 21.10.Dr \sep 21.10.Pc                   \sep 21.30.Fe
% 04.20.Jb Exact solutions
% 21.10.Gv Nucleon distributions and halo features
% 21.10.Ma Level density
% 21.10.Pc Single-particle levels and strength functions
% 21.10.-k 	Properties of nuclei; nuclear energy levels
% 21.10.Dr 	Binding energies and masses
% 21.10.Ma 	Level density
% 21.30.Fe Forces in hadronic systems and effective interactions
% 21.60.-n Nuclear structure models and methods
% 21.60.Cs Shell model
% 27.20.+n Properties of specific nuclei listed by mass ranges: 6 ≤ A ≤ 19
% 27.80.+w 	190 ≤ A ≤ 219( 	Properties of specific nuclei listed by mass ranges)
\end{keyword}
  
\end{frontmatter}

\section{Introduction} \label{sec.introduction}
The theoretical and experimental study of drip-line nuclei became increasingly important in the last forty years since the discovery of the Borromean Lithium isotope \cite{1985Tanihata}. From the point of view of basic knowledge, it is desirable to reach a deeper understanding of the properties of loosely bound nuclei by studying for example, the spectroscopy as well as the role of pairing in drip line nuclei. From application point of view, exotic nuclei are important, for example in astrophysical processes \cite{1993Kratz,2016Meisel} and as feeder to produce very heavy elements \cite{2010Oganessian}. A direct indication of the importance of this exotic nuclei is the fact that there are all around the world nuclear radioactive beam facilities which are increasing their mass and energy scope \cite{2012WhitePaper}.  This area of nuclear research will remain prominent for many years to come \cite{2015Longerp}.

The new ingredient of the exotic nuclei with respect to the nuclei along the stability valley is the continuum. This implies that any theoretical model which aim is to describe the properties of exotic nuclei must consider the coupling with the continuum spectrum of energy. Many of such theoretical models are described in references \cite{1969Mahaux,2001Grasso,2002IdBetan,2002Michel,2003Okolowicz,2005Bhattacharya,2006Volya,2009Michel,2012Hagen,2013Papadimitriou}.

One of the main ingredient to study the properties of nuclei is the pairing, `pairing lies at the heart of nuclear physics' \cite{2002Ria} and it is expected that this strong statement be true also for drip line nuclei \cite{1999prcBennaceur}. The pairing part of the particle-particle interaction increases the stability of nuclei close to beta-stability \cite{1964Lane}, likewise one may hope that the pairing in the continuum tilts the balance also towards the stability for nuclei close to the drip line \cite{2005Brink}. Recently, the treatment of the pairing in the Bardeen-Cooper-Schrieffer (BCS) approximation was extended by including the continuum through a basis of transformed harmonic oscillators \cite{2016Lay}.

As pairing carries much of the weight of the particle-particle interaction it is very handy for large scale mass calculation, where the use of effective interaction is computationally expensive due to the rapidity increase of dimensionality \cite{1995Aboussir}. In particular, the continuum pairing may be useful to calculate binding energies of heavy nuclei close to the drip line required in stellar nucleosynthesis calculations \cite{2007Arnould}. It is expected that the inclusion of the continuum pairing improve the binding energies theoretically estimated by extrapolating the known masses from the $\beta$ stability region \cite{1991Pearson,2001Goriely}.

In this work we solve the constant pairing Hamiltonian in the BCS \cite{1957Bardeen,1963Kisslinger,1964Lane} and in the Lipkin-Nogami (LN) \cite{1960Lipkin,1964Nogami,1966GoodFellow,1973Pradhan} approximations. The continuum spectrum of energy is incorporated through the continuum single particle level density (CSPLD) \cite{1997Sandulescu,2012npaIdBetan}. This approach has two advantages: first, the CSPLD modulates the pairing in the continuum, and second it saves a large amount of computing time connected with the model space size when continuum is considered. In particular, the state-independent large-scale mass-calculation  \cite{1987Tondeur} would benefit from this approach.

After this introductory section, the paper describes the formalism in section \ref{sec.formalism}. In section \ref{sec.applications}, a selected number of nuclear properties for the Tin chain, from the proton drip line to the neutron drip line, are calculated and compared with other formalisms and with experiment when they exist. In the final section \ref{sec.conclusions} we summarize our main finding and draw our conclusions.

%%%%%%%%%%%%%%%%%%%%%%%%%%%%%%%%%%%%%%%%%
%%%%%%%%%%%%%%%%%%%%%%%%%%%%%%%%%%%%%%%%%
\section{Formalism} \label{sec.formalism}
We are going to calculate some physical magnitude in a continuum energy representation using a constant pairing interaction in the BCS and LN approximations. The continuum single particle states are normalized to Dirac delta, then it results more simple to manipulate the equations in a box representation. After the equation have been obtained, we make the formal limit of the size of the spherical box to infinite.

\subsection{Hamiltonian} \label{sec.formalism.h}
The many-body system is described by the constant pairing interaction
\begin{equation}\label{eq.h}
 H = \sum_\alpha \varepsilon_a a_\alpha^\dag a_\alpha 
   -G P^\dag P \, ,
\end{equation}
where the index $\alpha = \{ a, m_\alpha \}=\{ n_a, l_a, j_a, m_{\alpha} \}$ label the single particle states. The mass-dependent strength $G$ is parametrized by the total number of particles, $A = A_{\textrm{core}} + A_{\textrm{valence}}$ and the relative neutron excess $I=\frac{N-Z}{A}$ as \cite{1969Nilsson,1988Madland}
\begin{equation}\label{eq.chi}
  G=\frac{\chi_{_1}}{A}(1-\chi_{_2} I) \, ,
\end{equation}
where $\chi_{_2}=0.385$ MeV \cite{1969Nilsson} and the constant $\chi_{_1}$ is adjusted to reproduce the experimental gap. 

The pair creation operator reads
\begin{equation}
 P^\dag = \sum_{\alpha > 0} a_\alpha^\dag a_{\bar{\alpha}}^\dag \, .
\end{equation}
The summation $\alpha > 0$ refers to the positive values of the projection of the total angular momentum $m_\alpha$, while the operator $a_{\bar{\alpha}}^\dag=(-1)^{j_a-m_\alpha} a_{a,-m_\alpha}^\dag$ is the time reverse of the $a_\alpha^\dag$ operator. The creation $a_\alpha^\dag$ and annihilation $a_\alpha$ operators creates bound and continuum states in a spherical box satisfying the usual anti-commutation relationship $\{ a_\alpha^\dag,a_{\alpha'} \} = \delta_{\alpha \alpha'}$. 

Even when negative and positive energy states are normalized in the box representation, the assignment of the same constant matrix elements of the interaction between particles in bound and continuum configurations is nonphysical \cite{1996Dobaczewski}. The influence of the non-resonant continuum is reduced when for the density energy is used the difference between the mean field and free density \cite{1937Beth}.

\subsection{Lipkin-Nogami Equations} \label{sec.formalism.ln}
The pairing interaction is diagonalized in the LN and the BCS approximations, with the LN Hamiltonian given by \cite{1960Lipkin,1964Nogami}
\begin{equation}
 H_{LN} = H - \lambda_1 N - \lambda_2 N^2 \, ,
\end{equation}
where $H$ is the original Hamiltonian (\ref{eq.h}) and $N=\sum_\alpha a_\alpha^\dag a_\alpha$ is the number operator. The introduction of the term $\lambda_2 N^2$ reduces the effect of the number fluctuation \cite{1960Lipkin,1964Nogami}, which is significant in the BCS approximation when the number of particles is small.

The LN equations in the box representation are given by \cite{1973Pradhan}
\begin{eqnarray}
 \frac{4}{G} &=& \sum_{n}{\frac{1}{E_n}} \label{eq.5} \\
 N &=& \sum_{n}{v_{n}^2} \label{eq.6} \\
 \frac{4\lambda_2}{G} &=& \frac{\left(\sum_{n}{u_{n}^{3}v_{n}}\right)\left(\sum_{n}{u_{n}v_{n}^{3}} \right) 
 - 2\sum_{n}{\left(u_{n}v_{n} \right)^{4}}}{\left(\sum_{n}{\left(u_{n}v_{n} \right)^{2}} \right)^{2} 
 - 2\sum_{n}{\left(u_{n}v_{n} \right)^{4}}} \, , \label{eq.7}
\end{eqnarray}
where the index $n=\{n_b, n_c\}$ labels the bound states $n_b$ and the box-continuum states $n_c$; representing the negative and positive energy states respectively, with
\begin{eqnarray}
 v_{n_b}^{2} & = & \frac{1}{2}\left( 1 - \frac{e_{n_b}}{E_{n_b}} \right)
   \quad, \quad 
 v_{n_b}^2 + u_{n_b}^2 = 1 \label{eq.v2} \\
 E_{n_b} & = & \sqrt{e_{n_b}^2 + \Delta^{2}} + \lambda_2 \label{eq.qp} \\
 e_{n_b} & = & \varepsilon_{n_b} - \lambda + (4\lambda_{2} - G)v_{n_b}^2 \\
 \lambda & = & \lambda_{1} + 2 \lambda_{2} (N+1)  \label{eq.l} 
\end{eqnarray}
and the same for the box-continuum states. The pairing gap reads,
\begin{equation}
  \Delta=\frac{G}{2}\sum_{n}{u_{n}v_{n}} \, . \label{eq.f} 
\end{equation}

We are now in condition to give the equations for the continuum representation. For this purpose we extend the size of the spherical box to infinite. In this limit the single particle density for the bound states is represented by $\sum_{n_b} \delta(\varepsilon - \varepsilon_{n_b})$. On the other hand, the box-continuum states become more dense and are represented by the continuum single-particle level density $g(\varepsilon)$ \cite{1937Beth}, with
\begin{equation}%\label{eq.ge}
 g(\varepsilon) = \frac{1}{\pi}\sum_{lj}{(2j+1)\frac{d\delta_{lj}}{d\varepsilon}} \, .
\end{equation}
The density for bound and continuum states can be written as
\begin{equation}
  \tilde{g}(\varepsilon) = \sum_{n_b} (2j_{n_b}+1) \delta(\varepsilon - \varepsilon_{n_b}) \theta(-\varepsilon) + g(\varepsilon) \theta(\varepsilon) \, .
\end{equation}

Magnitudes which in the box representation are calculated as $\sum_n f_n$, in the continuum representation take on the following form:
\begin{eqnarray}
  \sum_n f_n &\rightarrow&
     \sumint f \, ,
\end{eqnarray}
where
\begin{eqnarray}
     \sumint f &=& 
    \int\limits_{-\infty}^\infty \!\!\!   d\varepsilon
      \left[ \sum_{n_b} (2j_{n_b}+1) \delta(\varepsilon-\varepsilon_{n_b}) f(\varepsilon) \theta(-\varepsilon) \right.  \nonumber \\
    &&\left.  + 
       g(\varepsilon) f(\varepsilon) \theta(\varepsilon)
      \right]  \nonumber \\
    &=&
     \sum_{n_b} (2j_{n_b}+1) f_{n_b} 
       + \int\limits_0^\infty \!\! d\varepsilon \,  
                       g(\varepsilon) f(\varepsilon) \, .
\end{eqnarray}
The symbol $\sumint$ denotes both a summation over bound states and an integration over the continuous part of the energy spectrum. The integral is calculated using Gauss-Legendre quadrature. Hence, the CSPLD contribution seems to be as the natural extension of the contribution of the discrete part of the representation.

The BCS equations are obtained by taking $\lambda_2=0$ in the LN equations (\ref{eq.5})-(\ref{eq.f}) with $\lambda_1$ being the Fermi level. For this special case Eqs. (\ref{eq.5}) and (\ref{eq.6}) reduce to Eqs. (11) and (10) of Ref. \cite{1997Sandulescu} which include the CSPLD.

%%%%%%%%%%%%%%%%%%%%%%%%%%%%%%%%%%%%%%%%%
\subsection{Single particle representation. Bound and continuum states} \label{sec.formalism.mf}
The averaged single nucleon dynamics in the field of all other nucleons is used as a starting point in all many body methods. The single particle model space is calculated in a Woods-Saxon (WS) plus spin-orbit potential \cite{1954WS,1982Vertse},
\begin{eqnarray}\label{eq.mf}
  V_{WS}(r) &=& - \frac{V_0}{1+e^{\frac{r-R}{a}}} \, , \\
  V_{so}(r) &=& - \frac{V_{so}}{ra} \frac{2}{\hbar^2} \frac{e^{\frac{r-R}{a}}}{(1+e^{\frac{r-R}{a}})^2}\, .% \\
%   e         &=& e^{\frac{r-R}{a}} \, .
\end{eqnarray}

The mean field parameters, diffuseness $a$, the radius parameter $r_0$ in $R=r_0 A^{1/3}$ and the strengths $V_0$ and $V_{so}$ (considered mass-dependent \cite{1969Becchetti})
\begin{eqnarray}
   V_0    &=& c_0      - c_1\, I     \label{eq.mf1}  \\
   V_{so} &=& c_0^{so} + c_1^{so}\, I  \label{eq.mf2} \\
   I      &=& \frac{N-Z}{N+Z} \quad \textnormal{relative neutron excess}
\end{eqnarray}
are chosen to reproduce as well as possible the low lying experimental energies \cite{2007Schwierz} of the core plus one neutron. 

The single particle representation is formed by the valence bound and continuum states. The continuum part of the spectrum is represented by the CSPLD $g(\varepsilon)$, defined relative to the free particle density \cite{1937Beth,2005Charity}
\begin{equation}\label{eq.ge}
 g(\varepsilon) = \frac{1}{\pi}\sum_{lj}{(2j+1)\frac{d\delta_{lj}}{d\varepsilon}} \, ,
\end{equation}
where the phase shifts are calculated using the code of Ref. \cite{1995Ixaru}.
% 

% % 
% % % % % % % % % % % % % % % % % % % % %
\section{Applications} \label{sec.applications}
The numerical calculated gaps are compared with the following experimental five-point (fourth-order) gap equation \cite{2000Bender,2002Duguet}
\begin{eqnarray}\label{eq.d}
  \Delta(N) = \frac{1}{8}
               &&\left[
                 B(Z,N+2)-4B(Z,N+1)+6B(Z,N) \right. \nonumber \\
           &&     \left.  -4B(Z,N-1)+B(Z,N-2)
            \right] \, ,
\end{eqnarray}
where $B$ are the binding energies. The fourth-order gap equation has the advantage over the second- and third-order equations that the former is more precise than the second one and has not the problem of the ambiguity assignment of the last one. The experimental masses are taken from Ref. \cite{2003Audi}. In the case that the experimental masses are unknown, we use theoretical masses from Ref. \cite{2005Koura,2005Koura_html}.

\subsection{Single particle representation. Bound and continuum states}\label{sec.applications.mf}
On the proton drip-line side of the Tin isotopes, extrapolation from heavier isotopes would indicate that the ground state of the $^{101}$Sn nucleus would be the $5/2^+$ \cite{nndc} while the excited state would be $7/2^+$ \cite{2007Seweryniak}. Reference \cite{2010Darby} find ``strong experimental evidence'' that the reverse order actually occurs. In this paper we use $g_{7/2}$ as the ground state for the $^{101}$Sn. 

\begin{figure}[!ht]
\begin{center}
\vspace{6mm}
  \includegraphics[width=0.47\textwidth]{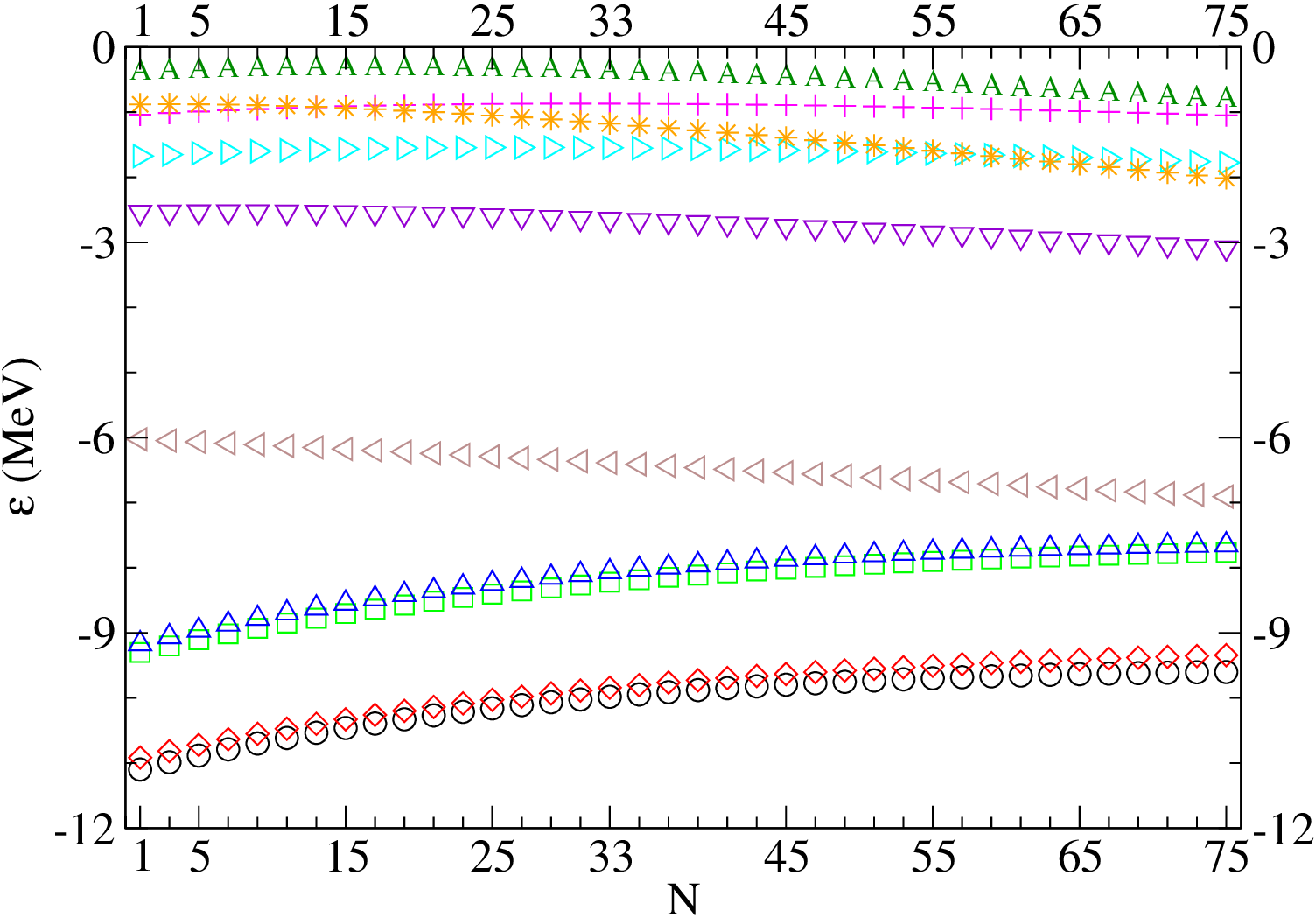}
  \caption{(color online) Evolution of the single particle energies in the core $^{100}$Sn as a function of the number of the valence neutrons. 
  The following labels \textcolor{black}{$\circ$}$g_{7/2}$, 
  \textcolor{red}{$\diamond$}$d_{5/2}$, 
  \textcolor{green}{$\square$}$s_{1/2}$, \textcolor{blue}{$\triangle$}$d_{3/2}$, \textcolor{brown}{$\lhd$}$h_{11/2}$, \textcolor{Fuchsia}{$\triangledown$}$f_{7/2}$, \textcolor{cyan}{$\rhd$}$p_{3/2}$, \textcolor{magenta}{$\plus$}$p_{1/2}$, \textcolor{orange}{$\ast$}$h_{9/2}$, {\scriptsize \textcolor{OliveGreen}{A}}$f_{5/2}$ 
  identify each single particle state.}
  \label{fig.evsn}
\vspace{6mm}
\end{center}
\end{figure}

We are going to calculate physical magnitudes for Tin isotopes from 102 up to 176. Since this is a huge mass interval, we decided to consider mass-dependent single particle energies (SPE). These SPE, shown in Fig. \ref{fig.evsn}, are calculated using the mean field parameters of Eq. (\ref{eq.mf1}) and (\ref{eq.mf2}) given in Table \ref{table.c}, being $^{100}$Sn the inert core. The parameters were fixed in order to reproduce as well as possible the low-lying neutrons energies of the $^{101}$Sn and $^{133}$Sn shown in Table \ref{table.esp}.

%%%%%%%%%%%%%%%%

\begin{table}[!ht]
\begin{center}
\caption{\label{table.c} Mean field parameters which define the mass dependent mean field strengths.}
\vspace{2mm}
% \begin{ruledtabular}
\begin{tabular}{ll}
\hline
   $c_0=51.95$ MeV        & $c_1=34.856$ MeV \\
   $c_0^{so}=11.30$ MeV\,fm & $c_1^{so}=9.075$ MeV\,fm \\ 
\hline

  $r_0=1.27$ fm & $a=0.7$ fm \\
\hline
\end{tabular}
% \end{ruledtabular}
\end{center}
\end{table}
%%%%%%%%%%%%%%%%

\begin{table}[!ht]
\begin{center}
  \caption{\label{table.esp} Low-lying neutron single particle energies $\varepsilon$ (MeV) in the $^{100}$Sn and $^{132}$Sn cores. The energy splitting between the ground state and the first excited state of $^{101}$Sn was taken from \cite{2007Seweryniak}, while the order was taken from \cite{2010Darby}. The separation energy of the $^{101}$Sn and $^{133}$Sn are from Ref. \cite{nndc}.}
\vspace{2mm}
% \begin{ruledtabular}
\begin{tabular}{cccc}
\hline

  core       & state       & $\varepsilon$ (MeV) & $\varepsilon_{\textrm{Exp}}$ (MeV) \\
\hline
  $^{100}$Sn & $0g_{7/2}$  & -11.100       & -11.100 \\
  $^{100}$Sn & $1d_{5/2}$  & -10.916       & -10.928 \\
\hline
  $^{132}$Sn & $1f_{7/2}$  &  -2.442       & -2.402   \\
  $^{132}$Sn & $2p_{3/2}$  &  -1.395       & -1.548  \\
\hline

\end{tabular}
\end{center}
% \end{ruledtabular}
\end{table}

\begin{figure}[!ht]
\begin{center}
\vspace{6mm}
  \includegraphics[width=0.45\textwidth]{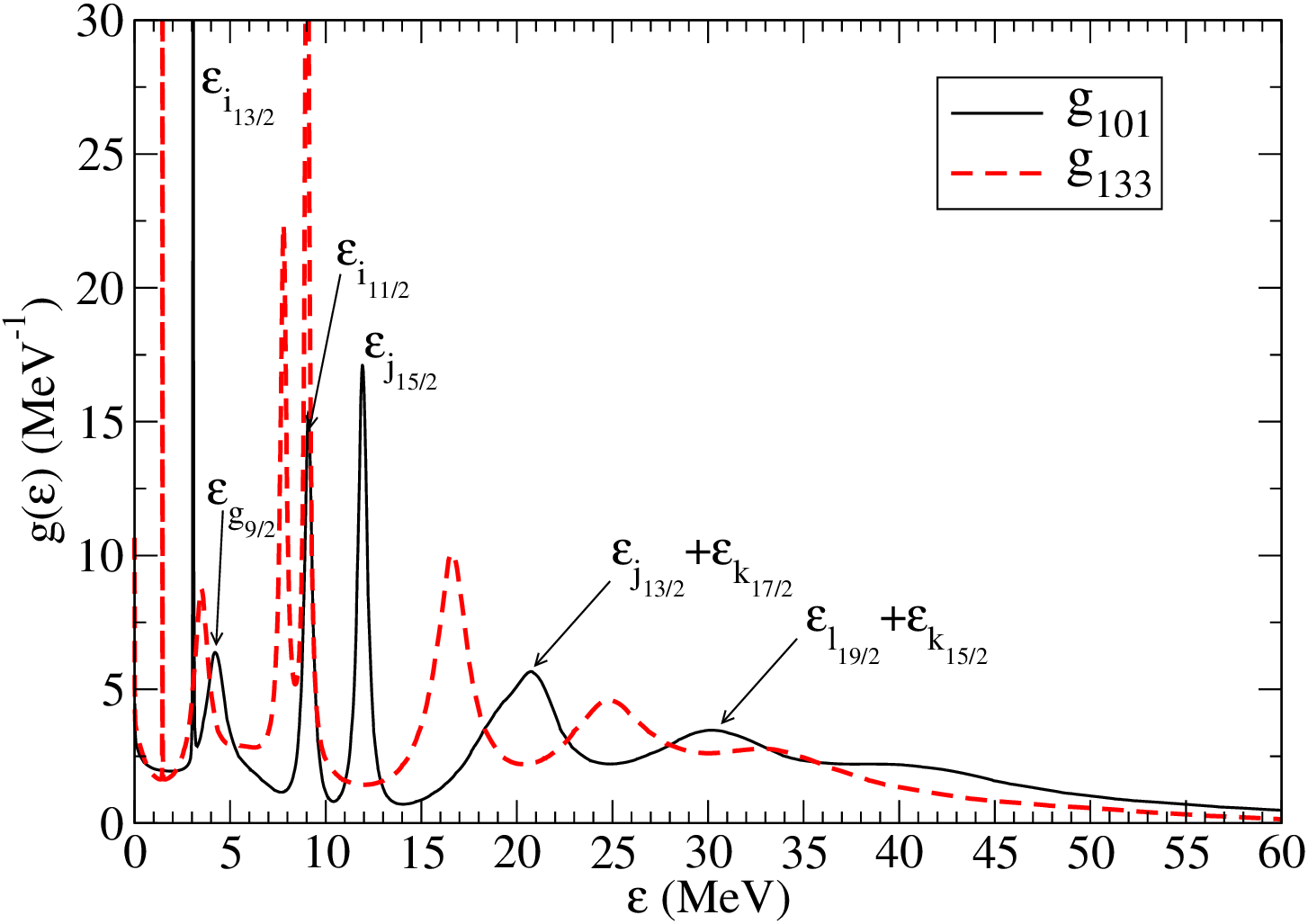}
  \caption{Continuum single particle level densities for a neutron in the cores $^{100}$Sn and $^{132}$Sn.}
  \label{fig.gvse}
\vspace{6mm}
\end{center}
\end{figure}

Figure \ref{fig.gvse} shows the CSPLD $g(\varepsilon)$ of $^{101}$Sn and $^{133}$Sn labeled as $g_{_{101}}$ and $g_{_{133}}$, respectively. These densities were calculated \cite{1995Ixaru} with angular momentum cut-off $l_{\textrm{max}}=10$ and energy cut-off $\varepsilon_{\textrm{max}}=60$ MeV. The imprint of the resonances shapes the CSPLD. Very narrow resonances appear at low energies whereas some superposition of wide resonances show up at higher energies. The labels $\varepsilon_{lj}$ and $\varepsilon_{lj}+\varepsilon_{l'j'}$ in this figure are used to identify the resonances and a superposition of two of them, respectively in the $^{101}$Sn nucleus. There is one to one association between the peaks of the CSPLD of the nuclei $^{101}$Sn and $^{133}$Sn. The overall structure of the density of the $^{133}$Sn remains invariant, but its peaks are shrink and displaced towards the continuum threshold.
In Ref. \cite{1994Nazarewicz} it was found that the high density of single particle states in the particle continuum produces an increase of BCS pairing correlations using the isospin dependent strength (\ref{eq.chi}). It will be shown that none of the magnitudes calculated using the CSPLD of Fig. \ref{fig.gvse}, nor even the pairing gap, show any unrealistic behavior.
%
%
%%%%%%%%%%%%%%%%%%%%%%%%%%%%%%%%%%%%%%%%%%%%
\subsection{Gap parameter}
In this section we are going to calculate the pairing gap in the BCS and LN approximations and we are going to compare the results with that obtained from the five-point gap equation (\ref{eq.d}).

\begin{figure}[!ht]
\begin{center}
\vspace{6mm}
  \includegraphics[width=0.45\textwidth]{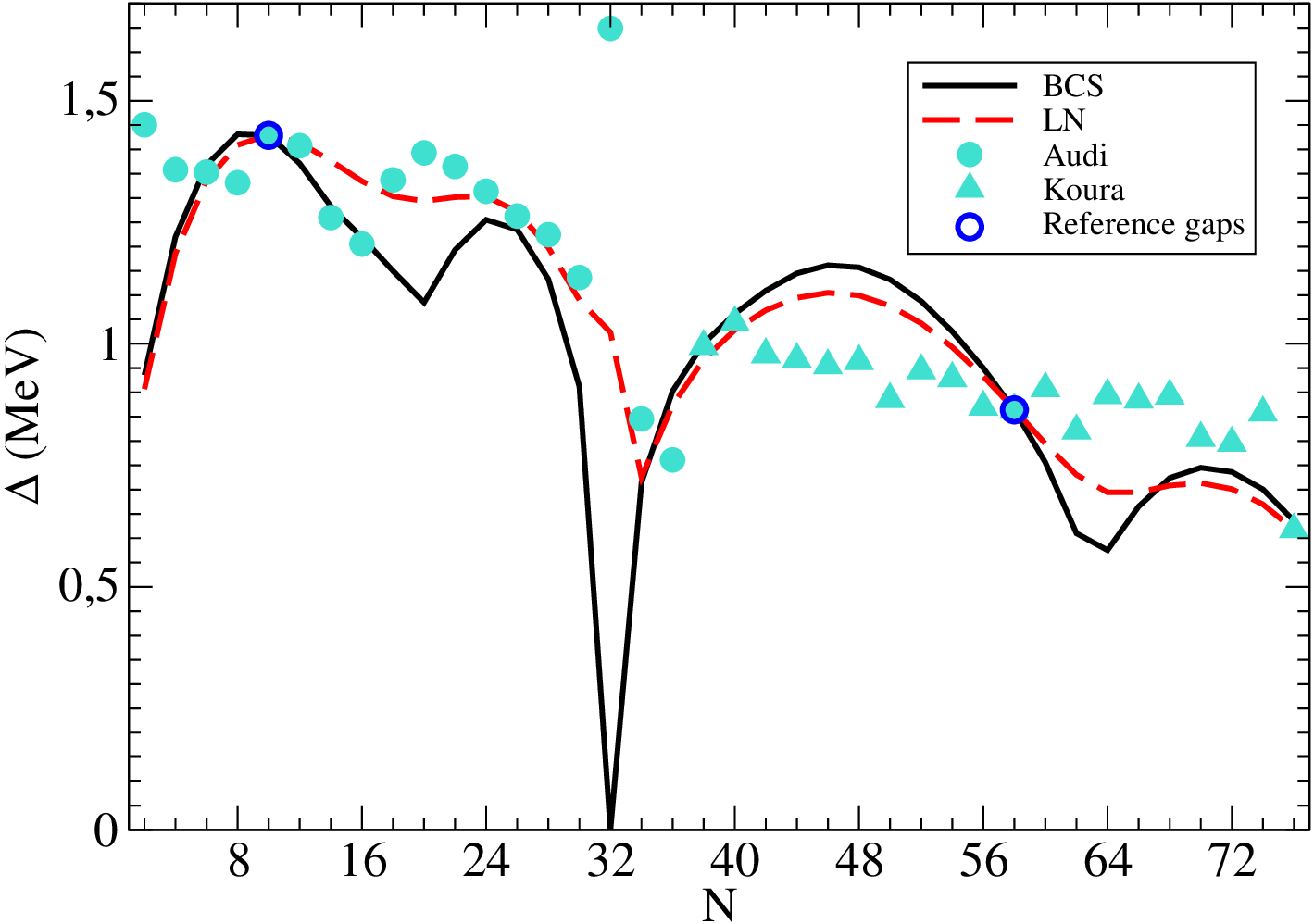}
  \caption{Comparison of the calculated gap in the BCS and LN approximations with the five-point gap equation (\ref{eq.d}). For $N< 36$ we use the experimental mass from Ref. \cite{2003Audi}, while for $N\ge36$ we use theoretical mass from Ref. \cite{2005Koura}.}
  \label{fig.gap}
\vspace{6mm}
\end{center}
\end{figure}

The parameter $\chi_{_1}$ of Eq. (\ref{eq.chi}) is different for each neutron major shells $50-82$ and $82-126$, and also for each approximate solution BCS or LN. The different $\chi_{_1}$ were chosen to reproduce the gaps of the isotopes $^{110}$Sn and $^{158}$Sn. These so called reference gaps were calculate using equation (\ref{eq.d}). For the isotope $^{110}$Sn the experimental masses of Ref. \cite{2003Audi} were used, while for the isotope $^{158}$Sn the theoretical mass of Ref. \cite{2005Koura}, were used. Table \ref{table.chi} gives the values of the different $\chi_{_1}$ for both reference gaps. For the first major shell we choose as reference isotope the $^{110}$Sn, because it is in the middle of the almost degenerate $g_{7/2}$ and $d_{5/2}$ shells. For the second major shell we found that the qualitative behavior shown in Fig. \ref{fig.gap} does not change with the election of different reference isotopes. Then, we choose as reference the isotope $^{158}$Sn because this election distributes evenly the deviation of the corresponding curves with respect to the uniform theoretical gap calculated using the five points equation.
\begin{table}[!ht]
\begin{center}
\caption{\label{table.chi}
Values of the parameter $\chi_{_1}$ of Eq. (\ref{eq.chi}) for the two major shells which determine the pairing strengths in the BCS and LN approximations. The gap parameters $\Delta$ were calculated using equation (\ref{eq.d}) with data from Refs. \cite{2003Audi,2005Koura}}
\vspace{2mm}
% \begin{ruledtabular}
\begin{tabular}{ccc}
\hline

  $\Delta$ (MeV)                       & $\chi_{_1}^{\rm{BCS}}$ (MeV)  & $\chi_{_1}^{\rm{LN}}$(MeV) \\
\hline
  $\Delta$($^{110}$Sn)=1.429 \cite{2003Audi}  & 18.637        & 18.269 \\
  $\Delta$($^{158}$Sn)=0.864 \cite{2005Koura} & 15.631        & 14.972 \\
\hline

\end{tabular}
\end{center}
\end{table}

Using the isospin pairing strength and the isospin single particle representation we calculate the pairing gap for the whole Tin isotope chain, from the proton drip line to the neutron drip line. Figure \ref{fig.gap} compares the five-points gap equation (\ref{eq.d}) with calculated gap in the BCS and LN approximations. 

As a general feature, the LN solution is smoother than the BCS one in the whole chain. In the first major shell the LN solution shows a better agreement with the experimental gap. The BCS values and the LN ones are more similar in the second major shell than in the first one. Besides, the LN result follows better the five-point gaps in the closure of the first major shell, except at $N=32$ where the five-point equation is not valid \cite{2000Bender}.

For the isotopes beyond $A=134$ the comparison between various theoretical mass tables \cite{1988Tachibana,1995Aboussir,2001Goriely,2005Koura} show significant differences between them. Thus, a comparison using the five points gap equation with theoretical mass would have little sense. In spite of this drawback, we decided to compare our results with the newest theoretical mass table in order to judge if our calculated gap in the neutron drip line side is reasonable. Our results for both, BCS and LN approximations, show the typical shell structure and they do not depart much from the overall behavior of the five points gaps.

The trend of our calculated gap fulfill the well known result for heavy nuclei \cite{1988Madland} that the average gap decreases towards the neutron-rich side and increases towards the proton-rich side. 

We compare our results with those of Ref. \cite{2000Bender} for the isotopes $^{134}$Sn-$^{164}$Sn. In this work the authors calculate the gap using the five points equation (\ref{eq.d}) from state-dependent BCS solution. For the comparison we used the result of their delta force model since it is smoother than the density-dependent delta interaction result. From Fig. 4 (upper) (of Ref. \cite{2000Bender}) we can see that the BCS gap lie in the band 0.65-1.1 MeV similar to our BCS 0.58-1.16 MeV. While from Fig. 7 (upper) (of Ref. \cite{2000Bender}) we can see that the LN gap lie in the band 0.55-1.3 MeV similar to our LN 0.7-1.1 MeV except in the vicinity of $^{164}$Sn.

A final observation from Fig. \ref{fig.gap} by considering the full range of $N$ is that the differences between the solutions of the BCS and LN are more pronounced far from the neutron drip line, i.e. the dependence of the gap on the model solution is less sensitive in the drip line region.

%%%%%%%%%%%%%%%%%%%%%%%%%%%%%%%%%%%
\subsection{Quasiparticle energies and occupation probability}\label{sec.qpe}
In this section we calculate the quasiparticle energies Eq. (\ref{eq.qp}) in the BCS and LN approximations and show the occupation probabilities Eq. (\ref{eq.v2}) for some selected isotopes. 

\begin{figure}[!ht]
\begin{center}
\vspace{6mm}
  \includegraphics[width=0.47\textwidth]{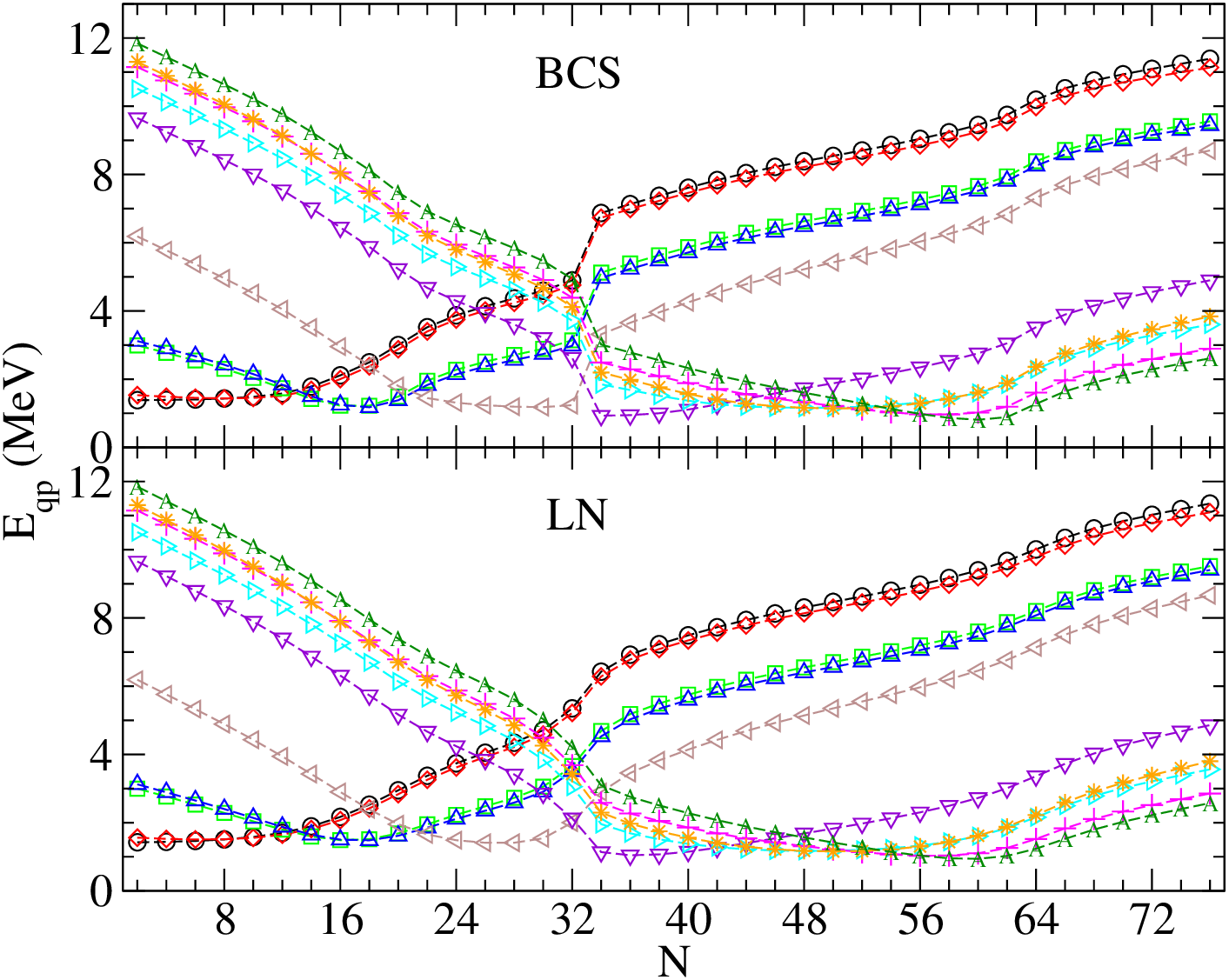}
  \caption{(color online) Evolution of the quasiparticle energies in the BCS (upper) and LN (lower) approximations as a function of the number of the valence neutrons in the $^{100}$Sn. 
  The particle states are identify with the following labels: \textcolor{black}{$\circ$}$g_{7/2}$, 
  \textcolor{red}{$\diamond$}$d_{5/2}$, \textcolor{green}{$\square$}$s_{1/2}$, \textcolor{blue}{$\triangle$}$d_{3/2}$, \textcolor{brown}{$\lhd$}$h_{11/2}$, \textcolor{Fuchsia}{$\triangledown$}$f_{7/2}$, \textcolor{cyan}{$\rhd$}$p_{3/2}$, \textcolor{magenta}{$\plus$}$p_{1/2}$, \textcolor{orange}{$\ast$}$h_{9/2}$, {\scriptsize \textcolor{OliveGreen}{A}}$f_{5/2}$.}
  \label{fig.eqvsn}
\vspace{6mm}
\end{center}
\end{figure}

Figure \ref{fig.eqvsn} shows the evolution of the quasiparticle energies as a function of the valence neutrons from $N=2$ to $N=76$. While the bound states accommodate up to sixty two neutrons, the continuum states, through the CSPLD, provide the needed configurations for the other fourteen nucleons. The order of the lowest states, $g_{7/2},d_{5/2},s_{1/2},d_{3/2},h_{11/2}$ in the first major shell, is like that of Ref. \cite{1996Dobaczewski} but the $g_{7/2}$ state, which appears in the fourth order. The lowest states in the second major shell are $f_{7/2},p_{3/2},h_{9/2},p_{1/2},f_{5/2}$ while in Ref. \cite{1996Dobaczewski} the order is $f_{7/2},p_{3/2},p_{1/2},f_{5/2},h_{9/2},i_{13/2}$. The last state is absent in Fig. \ref{fig.eqvsn} because it is a continuum state included in the CSPLD.

The gap between the two major shells decreases as $N$ increases, as can be seen from Fig. \ref{fig.eqvsn}, until the first major shell is completed at $N=32$ where the nucleus becomes normal. Both approximations (BCS and LN) show the crossing at the same isotope $^{132}$Sn, but in the LN approximation the crossing is smoother. Beside this issue there is not any other appreciable differences between the solution of both approximations.

It is known that the ground state of the nuclei $^{103}$Sn \cite{2001Fahlander} is the neutron single particle $d_{5/2}$ state. From Fig. \ref{fig.eqvsn} we observe that the crossing between the $g_{7/2}$ and $d_{5/2}$ levels happens at $^{109}$Sn, instead.

%%%%%%%%%%%%%%

In Table \ref{table.v2} we show some selected occupation probabilities. As a general feature the occupation increases with the number of particle until the first major shell is completed at $^{132}$Sn. Then, the states in the second major shell starting to be monotonically populated up to the nucleus $^{162}$Sn. Henceforth, the particles have no more bound configurations to fill and them have to populate the continuum spectrum of energy (this will be better analyzed in the next section, see Fig. \ref{fig.nvsn}). This feature is shown in Table \ref{table.v2} by the almost constant value of the bound-configuration occupations for particles greater than sixty two.
\begin{table*}
\begin{center}
\caption{\label{table.v2}
 Occupation probability for some selected Sn isotopes in the BCS and LN approximations.}
\vspace{2mm}
% \begin{ruledtabular}
{\tiny
\begin{tabular}{l|llllllllll}
\hline

       & \textrm{$g_{7/2}$}  & \textrm{$d_{5/2}$} & \textrm{$s_{1/2}$} & \textrm{$d_{3/2}$} & \textrm{$h_{11/2}$} 
                 & \textrm{$f_{7/2}$}  & \textrm{$p_{3/2}$} & \textrm{$p_{1/2}$} & \textrm{$h_{9/2}$} & \textrm{$f_{5/2}$}\\
  \hline
  \vspace{-0.16cm}     &   &  &  &  &  &  &  & &  &\\
  $^{102}$Sn(BCS)            & 0.131      & 0.104     & 0.025     & 0.023     & 0.006
                 & 0.002      & 0.002     & 0.002     & 0.002     & 0.002  \\
  $^{102}$Sn(LN)             & 0.260      & 0.215     & 0.051     & 0.046     & 0.011
                 & 0.004      & 0.004     & 0.003     & 0.003     & 0.003 \\
  \vspace{-0.2cm}     &   &  &  &  &  &  &  & &  &\\
  $^{110}$Sn(BCS)            & 0.633      & 0.582     & 0.144     & 0.127     & 0.026
                 & 0.008      & 0.006     & 0.006     & 0.006     & 0.005   \\
  $^{110}$Sn(LN)             & 0.574      & 0.619     & 0.164     & 0.145     & 0.028
                 & 0.008      & 0.007     & 0.006     & 0.006     & 0.005 \\
  \vspace{-0.2cm}     &   &  &  &  &  &  &  & &  &\\
  $^{120}$Sn(BCS)            & 0.966      &  0.963    & 0.843      & 0.809    & 0.101
                 & 0.011      & 0.008     & 0.006      & 0.006    & 0.005   \\
  $^{120}$Sn(LN)             & 0.943      & 0.938     & 0.764      & 0.728    & 0.154
                 & 0.017      & 0.012     & 0.010      & 0.010    & 0.008 \\
  \vspace{-0.2cm}     &   &  &  &  &  &  &  & &  &\\
  $^{124}$Sn(BCS)            & 0.973      & 0.971      &  0.917      &  0.905      &  0.358
                 & 0.022      & 0.014      &  0.011      &  0.012      &  0.009 \\
  $^{124}$Sn(LN)             & 0.966      & 0.964      & 0.891      & 0.875      & 0.373  
                 & 0.026      & 0.017      & 0.013      & 0.014      & 0.011   \\
  \vspace{-0.2cm}     &   &  &  &  &  &  &  & &  &\\
  $^{132}$Sn(BCS)            & 1.0        & 1.0       & 1.0       & 1.0       & 1.0
                 & 0.0        & 0.0       & 0.0       & 0.0       & 0.0   \\
  $^{132}$Sn(LN)             & 0.990      & 0.989     & 0.977     & 0.975     & 0.904
                 & 0.082      & 0.035     & 0.023     & 0.027     & 0.017 \\
  \vspace{-0.2cm}     &   &  &  &  &  &  &  & &  &\\
  $^{146}$Sn(BCS)            & 0.995       & 0.995       & 0.992       & 0.991       & 0.986  
                 & 0.840       & 0.445       & 0.206       & 0.364       & 0.125 \\
  $^{146}$Sn(LN)             & 0.995       & 0.995       & 0.992       & 0.992       & 0.987 
                 & 0.830       & 0.445       & 0.214       & 0.369       & 0.129 \\
  \vspace{-0.2cm}     &   &  &  &  &  &  &  & &  &\\
  $^{162}$Sn(BCS)           & 0.999      & 0.999      & 0.999      & 0.998      & 0.998
                 & 0.990      & 0.972      & 0.930      & 0.973      & 0.863 \\
  $^{162}$Sn(LN)             & 0.999      & 0.998      & 0.998      & 0.998      & 0.997
                 & 0.984      & 0.953      & 0.884      & 0.955      & 0.797 \\
  \vspace{-0.2cm}     &   &  &  &  &  &  &  & &  &\\
  $^{170}$Sn(BCS)           & 0.999      & 0.999      & 0.998      & 0.998      & 0.998
                 & 0.993      & 0.985      & 0.976      & 0.987      & 0.968 \\
  $^{170}$Sn(LN)             & 0.999      & 0.999      & 0.998      & 0.998      & 0.998
                 & 0.993      & 0.985      & 0.974      & 0.987      & 0.965 \\
\hline
\end{tabular}
}%end small
\end{center}
% }%end small
% \end{ruledtabular}
\end{table*}

%%%%%%%%%%%%%%%%%%%%%%%%%%%%%%%%%%%
\subsection{Fermi level and particle number}\label{sec.fl}
The Fermi level defined in Eq. (\ref{eq.l}) and particle number from Eq. (\ref{eq.6}) are calculated and discussed in this section. Both magnitudes are calculated in the BCS and LN approximations for the whole isotopic chain. 

\begin{figure}[!ht]
\begin{center}
\vspace{6mm}
  \includegraphics[width=0.47\textwidth]{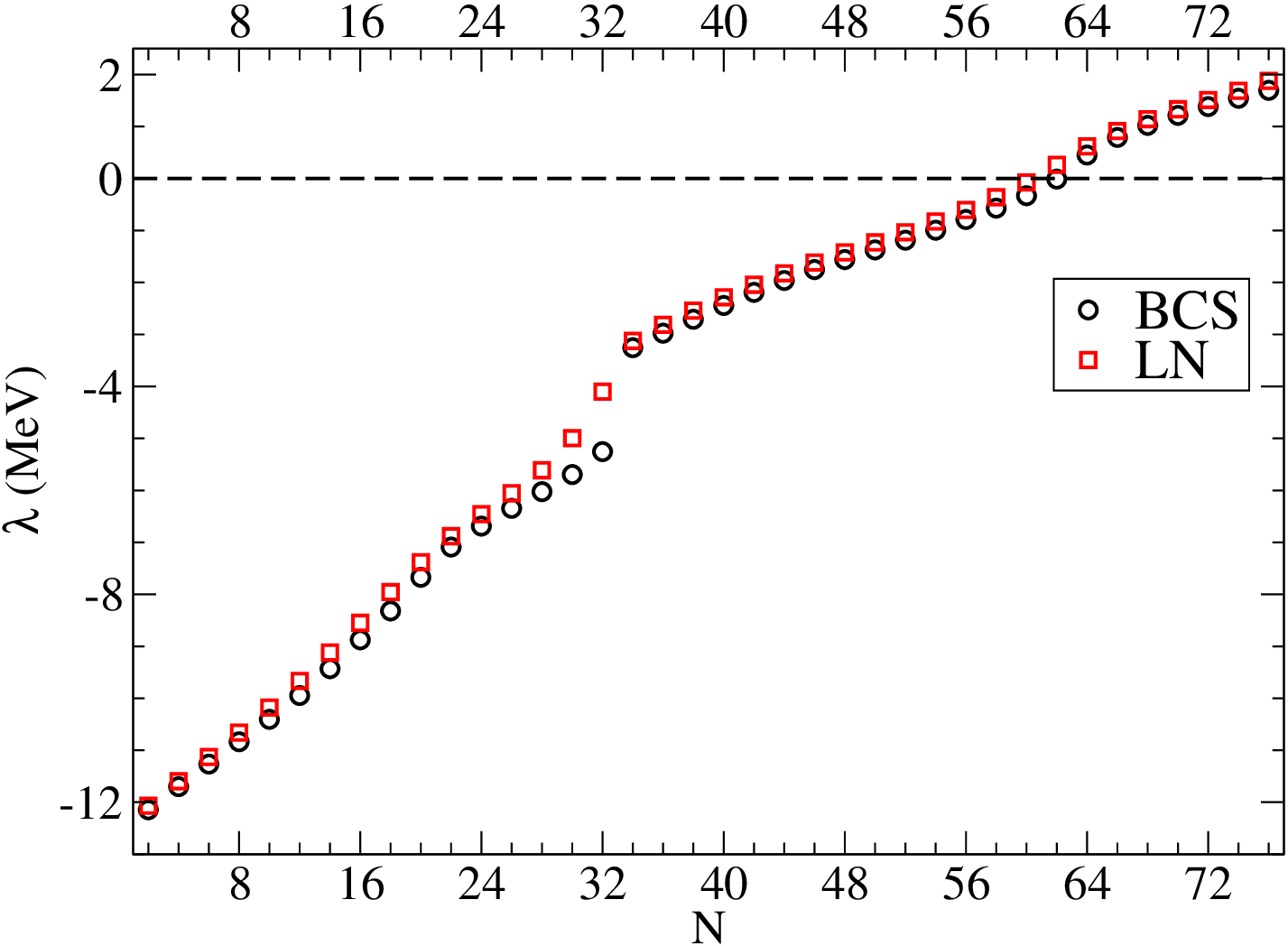}
  \caption{Evolution of the Fermi level in the BCS and LN approximations as a function of the number of the valence neutrons $N$ in the core $^{100}$Sn.}
  \label{fig.lvsn}
\vspace{6mm}
\end{center}
\end{figure}

The Fermi level, $\lambda(N)$, is shown in Fig. \ref{fig.lvsn} as a function of even number of valence neutrons. It has the usual linear dependence with the particle number for the whole isotopic chain. At $N=32$ the figure shows a change in the slope between the first and the second major shells, with smaller slope in the second one. The transition between these two linear behaviors is smoother in the LN approximation. After all bound states configurations have been filled up, the Fermi level crosses the continuum threshold at $N=62$. Like for the previously calculated magnitudes both BCS and LN approximations give very similar results.
%%%%%%%%%%%%

\begin{figure}[!ht]
\begin{center}
\vspace{6mm}
  \includegraphics[width=0.47\textwidth]{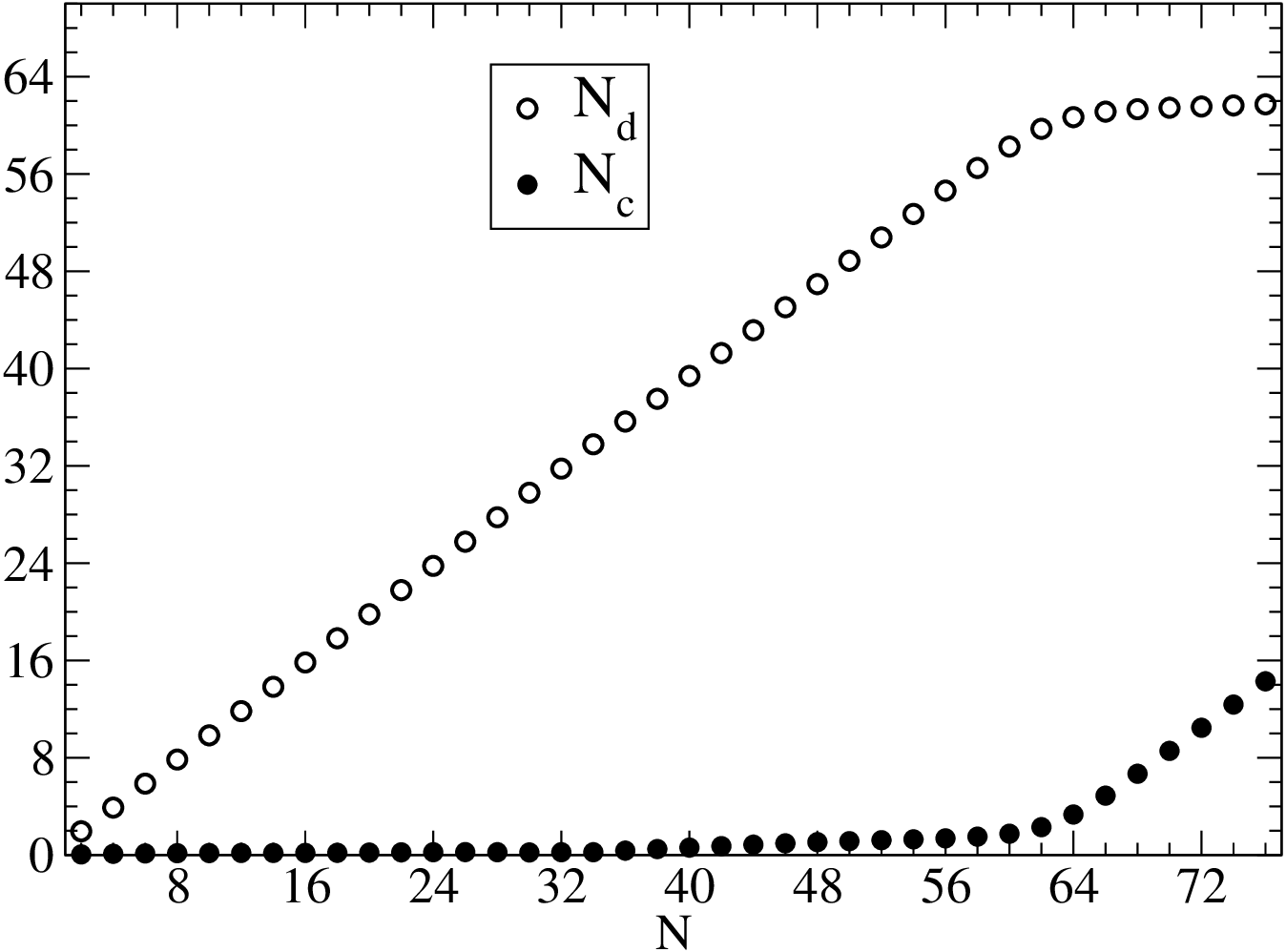}
  \caption{Discrete and continuum particle number as defined in Eq. (\ref{eq.nc}) in the LN approximation as a function of the number of the valence neutrons $N$. At this scale, the BCS solutions (not shown in the figure) would be almost indistinguishable of the LN one.}
  \label{fig.nvsn}
\vspace{6mm}
\end{center}
\end{figure}

In order to illustrate the contribution coming from the energy continuum spectrum we show in Fig. \ref{fig.nvsn} the discrete $N_d$ and the continuum $N_c$ particle number. They are defined from the bound and continuum contribution parts of the single particle representation, respectively as \cite{2012npaIdBetan},
\begin{equation} \label{eq.nc}
  N = N_b + N_c = \sum^{l_{\textrm{max}}}_{lj}\, (2j+1) v^2_{lj}
                + \int_0^{\varepsilon_{\textrm{max}}} v^2(\varepsilon) g(\varepsilon)\, d\varepsilon
\end{equation}
with $l_{\textrm{max}}=10$ and $\varepsilon_{\textrm{max}}=60$ MeV as defined in the section \ref{sec.applications.mf}.
%%%%%%%%%%%%
%%%%%%%%%%%%

Figure \ref{fig.nvsn} shows how the number of particle in the continuum $N_c$ remains small for all isotope up to $^{162}$Sn, i.e. as long as there is bound state available the particles will not populate the continuum. From $^{164}$Sn on the bound configurations are almost occupied, as can be seen in Table \ref{table.v2}, and $N_c$ starts to increases while $N_d$ approaches to $62$. This is an indication

% % % % % % % % % % % % % % % % % % % % % % % % 
\subsection{Binding energy and one- and two-neutron separation energies}
In this section we calculate the binding energy per nucleon, the two-neutron separation energy and the one-neutron separation energy in the non-blocking approximation. The results are compared with that of Refs. \cite{1996Dobaczewski,2013Isakov} and with experimental data \cite{2012Wang}. 

\subsubsection{Binding energy:} The binding energy per nucleon in the BCS and LN approximations are calculated from
\begin{eqnarray}
 \frac{B}{A} &=& \frac{ B(^{100}\textrm{Sn})-E_{\textrm{BCS/LN}}(N)}{50+N} \, ,
\end{eqnarray}
where the experimental value $B(^{100}\textrm{Sn})/100=8.253$ MeV \cite{2012Wang} was used for the core binding energy, and
\begin{eqnarray}
  E_{\textrm{BCS/LN}}(N)
 &=& \sumint_{n} v^2_n \left( \varepsilon_n - \frac{G}{2}\, v^2_n \right)
            - \frac{\Delta^2}{G}
            - \lambda_2 \sumint_{n} 2\, u^2_n\, v^2_n \, . \nonumber \\
\end{eqnarray}

\begin{figure}[!ht]
\begin{center}
\vspace{6mm}
  \includegraphics[width=0.47\textwidth]{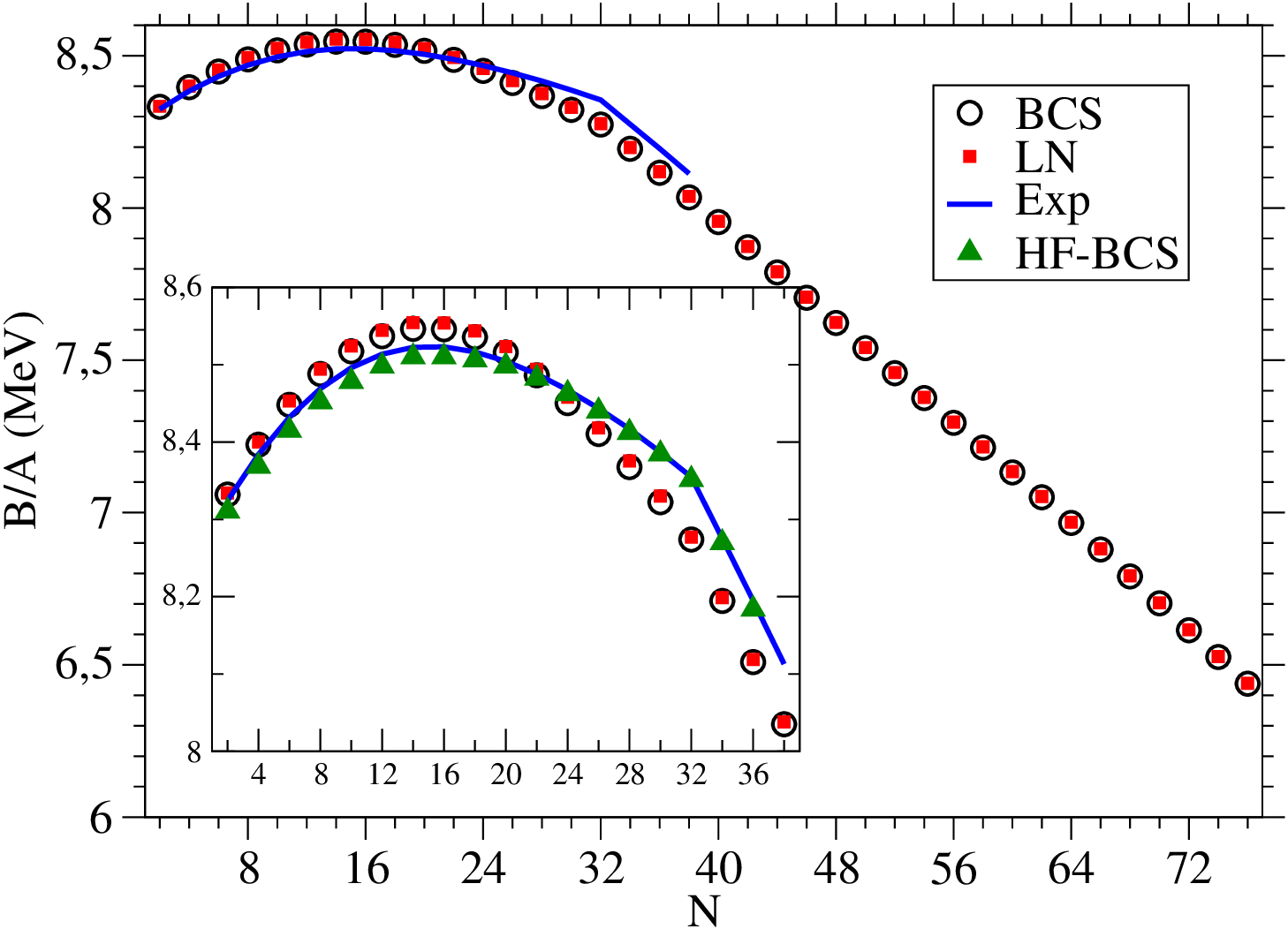}
  \caption{Numerical calculation of the binding energy per particle for the Sn isotopes from the BCS, LN and HF+BCS approximations. The numerical result of HF-BCS approximation is from Ref. \cite{2013Isakov}. The experimental data (Exp) are from the Atomic Mass Evaluation \cite{2012Wang}.}
  \label{fig.evsn2}
\vspace{6mm}
\end{center}
\end{figure}

Figure \ref{fig.evsn2} shows the calculated binding energy per particle versus the number of valence neutrons. Both BCS and LN approximations give similar results for all isotopes. The maximum binding energy per nucleon occurs at $N\sim14,16$ and coincides with the experimental maximum. At $N=32$ the figure shows a break in the experimental values. From here on, the numerical solutions have a linear behavior which follows the extrapolated slope of the experimental data. The insert in Fig. \ref{fig.evsn2} shows the binding energy per nucleon in the range $2 \le N \le 36$. The numerical solutions string along with the experimental values for $N\le 24$, while for $28 \le N \le 36$ our results depart from the experimental ones. The comparison with the result from the HF+BCS approximation of Ref. \cite{2013Isakov} shows that our precision is similar to this one up to $N=24$. Beyond this nucleus the HF+BCS perfectly agrees with experiment. The CSPLD gives and alternative representation to include the continuum single particle configurations to that of the spherical box \cite{1996Dobaczewski}. The inclusion of the continuum allows to calculate magnitudes in nuclei with extreme neutron-to-proton ratios and, in this way gives some insight about the behavior of these exotic nuclei beyond the present experimental data.

\subsubsection{Two-neutron separation energy:} 
\begin{figure}[!ht]
\begin{center}
\vspace{6mm}
  \includegraphics[width=0.47\textwidth]{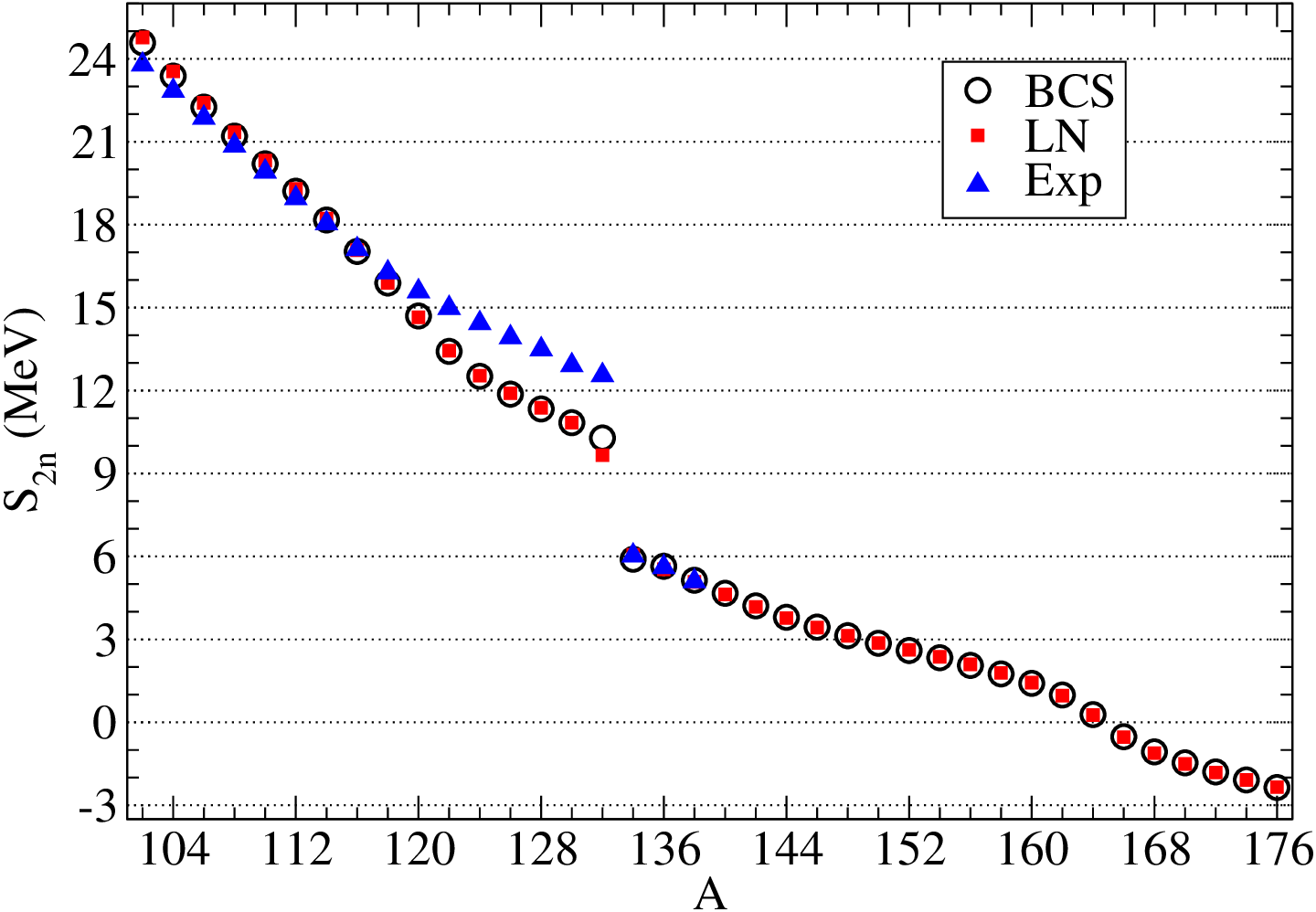}
  \caption{Two neutron separation energy for the Sn isotopes. The experimental data (Exp) are from the Atomic Mass Evaluation \cite{2012Wang}.}
  \label{fig.s2n}
\vspace{6mm}
\end{center}
\end{figure}

The two neutron separation energy is calculated from
\begin{equation}
  S_{2n}(N) = -[E_{\textrm{BCS/LN}}(N)- E_{\textrm{BCS/LN}}(N-2)] \, .
\end{equation}
Figure \ref{fig.s2n} shows the calculated two neutron separation energies and the experimental values \cite{2012Wang} as a function of the atomic mass $A=100+N$. We see a good agreement with experimental data up to $N=18$. From here on, both BCS and LN calculations depart from the experimental data reaching a maximum of around 2 MeV at $N=32$. Besides this awkward behavior at the closure of the first major shell, the matching with experimental data for the nuclei $^{134}$Sn to $^{138}$Sn is excellent. As stated before, the inclusion of the continuum would allow us to guess what to expect for observables close to the drip line.

In Ref. \cite{1994Nazarewicz} the continuum is included via harmonic oscillator (HO) states. Using BCS approximation, the authors get the two-neutron drip line at the nucleus $^{162}$Sn considering 12 HO shells in the single particle representation and at $^{166}$Sn using 40 HO shells. The more elaborated Hartree-Fock-Bogoliubov approximation gives that the $S_{2n}$ lies between $^{168}$Sn-$^{176}$Sn, depending on the interaction used. Our calculation shows that the two-neutron drip line occurs at $^{164}$Sn.

From Ref. \cite{1975Beiner} we know that the two-neutron separation energy behaves approximately as $S_{2n}(N) \approx -2\lambda(N-1)$. In the non-blocking approximation we take $\lambda(N-1)$ as the Fermi level of the even $N$ nucleus. Then, the Fermi level should change sign at the same nucleus as the two-neutron separation energy, which is approximately verified as can be seen from Figs. \ref{fig.lvsn} and \ref{fig.s2n}.

\subsubsection{One-neutron separation energy:} In the non-blocking approximation the single neutron separation energy is given approximately by \cite{1975Beiner}
\begin{equation}\label{eq.sn}
  S_n(N-1) = -\lambda(N)
      + \frac{1}{2} \left[  \frac{\lambda(N)-\lambda(N-2)}{2} \right]
      - E_{\textrm{min}}(N) \, ,
\end{equation}
where $\lambda$ was calculated in section \ref{sec.fl} and $E_{\textrm{min}}$ is the smallest of the quasiparticle energies as calculated in section \ref{sec.qpe}. Both of this quantities are calculated for even-$N$ isotopes. The approximation (\ref{eq.sn}) can not be applied for the isotopes $^{101}$Sn and $^{133}$Sn \cite{1975Beiner}.

\begin{figure}[!ht]
\begin{center}
\vspace{6mm}
  \includegraphics[width=0.47\textwidth]{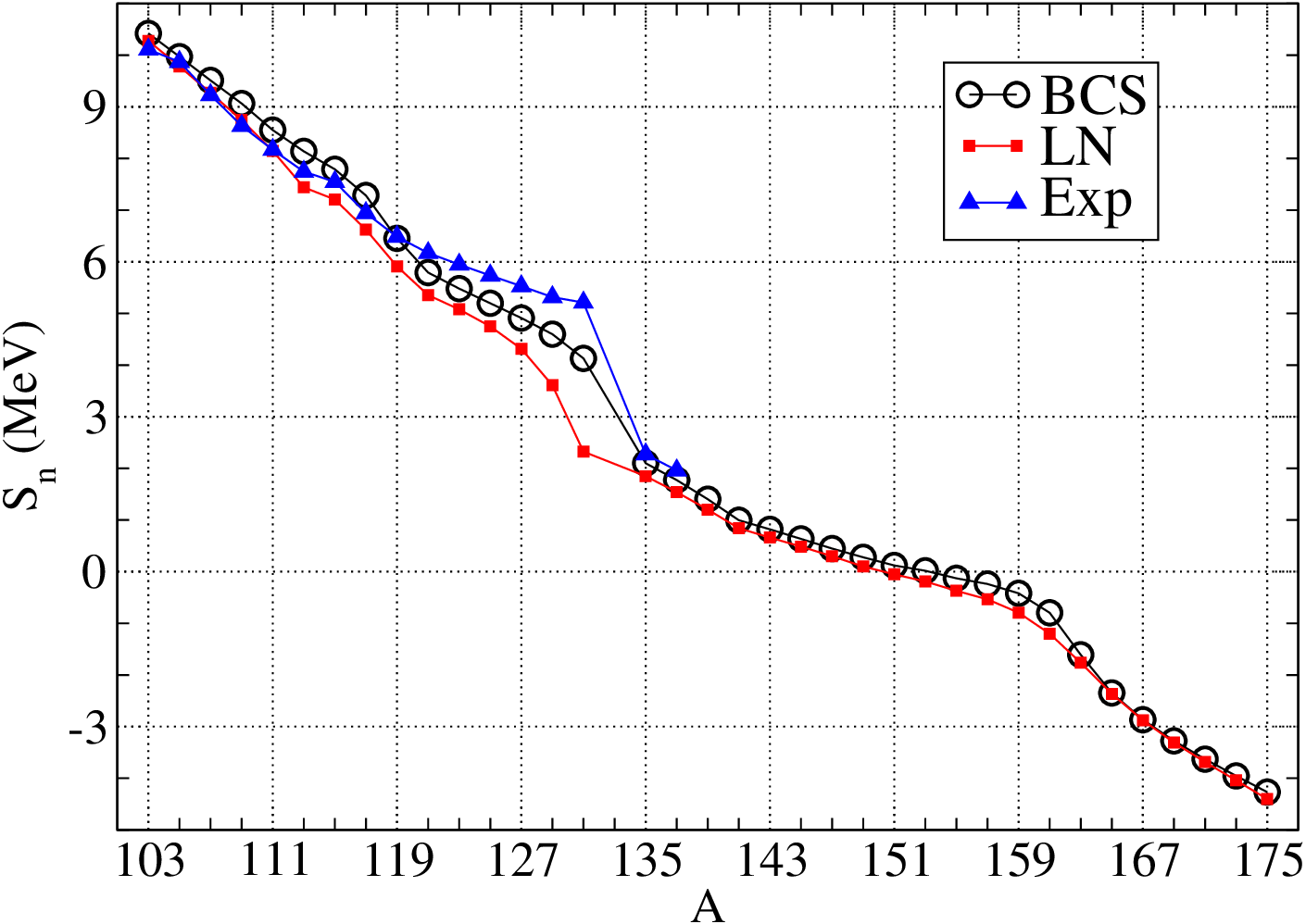}
  \caption{Neutron separation energies as a function of the atomic mass for the Tin isotopes for BCS and LN approximations. The experimental data (Exp) are from the Atomic Mass Evaluation \cite{2012Wang}.}
  \label{fig.sn}
\vspace{6mm}
\end{center}
\end{figure}

Figure \ref{fig.sn} compares the separation energy calculated in the BCS and LN approximations with that of experimental data from Ref. \cite{2012Wang}. The experimental separation energy for the isotopes of the first major shell decreases monotonically up to $A=119$ and then the slope decreases slightly. The numerical results follow the experimental data up to $A=119$ and then they match again at the beginning of the second major shell. 

Defining the drip line as the even nucleus at which $S_n$ (\ref{eq.sn}) changes sign between its odd neighbors, the BCS and LN approximations give the nuclei $^{154}$Sn and $^{150}$Sn, respectively. These results are close to that of Ref. \cite{1994Nazarewicz} which places the drip line at $^{148}$Sn.

% % % % % % % % % % % % % % % % % % % % % % 
% % % % % % % % % % % % % % % % % % % % % % 
\section{Discussion and conclusions} \label{sec.conclusions}
In this work, we have calculated several physical properties of the whole chain of the Tin isotope from the proton to the neutron drip line. For this purpose, the continuum spectrum of energy was included explicitly via the single particle level density. The properties calculated were the gap parameter, the Fermi level, the binding energy, and one- and two-neutron separation energies. Even when the proposed formulation to deal with the continuum spectrum is able to produce results beyond the drip line they may have not physical meaning. 

The solution of the pairing Hamiltonian were work out in the BCS and LN approximations. Due to the significant change of the proton-to-neutron ratio (from 0.96 in the proton drip line side to 0.66 in the neutron drip line side) the mean field, which defines the single particle representation, and the pairing strength were both considered isospin dependent.

In calculating the various magnitudes, both BCS and LN approaches, gave similar results. As an exception, when calculating the gap it can be seen that the LN approximation agrees quite better with experimental data than BCS, especially in the first major shell. As a distinctive feature of the comparison between the solutions of these two approaches, the variation of the physical magnitudes as a function of the valence neutron, always displayed a smoother behavior in the LN than the BCS solution. 

The gap parameter calculated with the BCS and LN approximations show the usual shell structure and follow in general the behavior of the experimental values. The gaps in the second major shell are smaller than the one in the first major shell, this is the expected tendency for the gap, i.e. it decreases with increasing number of neutrons. The Fermi level shows the typical linear dependence with $N$ for the whole chain of Tin isotopes considered. It crosses the continuum threshold at $N=62$, in which also changes the sign of the two neutron separation energy. The calculated binding energy shows that the maximum value occurs approximately at $N=14$, coinciding with experimental data. The comparison with the binding energy calculated within the HF+BCS approach of Ref. \cite{2013Isakov} (Fig. 1), shows a similar precision up to $N=24$. Finally, the calculated separation energies $S_n$ and $S_{2n}$, agree with the experimental values for the isotopes $A$ from 102 to 119 of the first major shell and $A$ between 134 and 138 of the second major shell.

In this first stage of the approach presented here, i.e. the use of the CSPLD with constant pairing, configures an economical way to include explicitly the continuum spectrum of energy in large scale mass calculation where the continuum boosts to the sky the computing time and memory requirement. In spite to the simplicity of this approach, the computed properties were in general in good agreement with more sophisticated approaches and with experimental results. There is in progress, an improvement of the presented method using state-dependent pairing interaction to overcome the founded discrepancies with experimental data.

% 
% 
% 
% % %%%%%%%%%%%%%%%%%%%%%%%
% % % \begin{acknowledgments}
\ack
%This work was supported  by the National Council of Research PIP-0625 (CONICET, Argentina).
% % % \end{acknowledgments}
% % % % % % % % % % % % % 
% % \begin{acknowledgments}
 This work has been supported by the Consejo Nacional de Investigaciones Cient\'{\i}ficas y T\'ecnicas PIP-625 (CONICET, Argentina).
% % \end{acknowledgments}
% % 
% % 
% % 
% % 
% % 
% % %%%%%%%%%%%%%%%%%%%%%%%%%%%%%%%%%%

%-------------
% Bibliography
%\bibliographystyle{elsarticle-num} 
%\bibliography{bcs2015}

\end{document}